\documentclass[preprint,prd,noshowpacs]{revtex4}
\usepackage{amssymb}
\usepackage{amsmath}
\usepackage{amsfonts}
\usepackage{graphicx, float}

\begin{document}

\title{Neutrino induced decoherence and variation in nuclear decay rates}

\author{Douglas Singleton}
\email{dougs@csufresno.edu}
\affiliation{Department of Physics, California State University, Fresno, CA 93740-8031 USA}
\author{Nader Inan}
\email{ninan@ucmerced.edu}
\author{Raymond Y. Chiao}
\email{rchiao@ucmerced.edu}
\affiliation{University of California, Merced, Schools
of Natural Sciences and Engineering, P.O. Box 2039, Merced, CA 95344, USA}

\date{\today}

\begin{abstract}
Recent work has proposed that the interaction between ordinary matter and a stochastic gravitational
background can lead to the decoherence of large aggregates of ordinary matter. In this work we point out
that these arguments can be carried over to a stochastic neutrino background but with the Planck scale of the gravitational
decoherence replaced by the weak scale. This implies that it might be possible to observe such neutrino
induced decoherence on a small, microscopic system rather than a macroscopic system as is the case for
gravitationally induced decoherence. In particular we suggest that neutrino decoherence could be
linked with observed variations in the decay rates of certain nuclei. 
Finally we point out that this proposed neutrino induced decoherence can be considered the complement of the 
Mikheev-Smirnov-Wolfenstein (MSW) effect.          
\end{abstract}

\maketitle

\section{Introduction}

In a recent paper \cite{blencowe}, it was pointed out that the interaction of matter
with the stochastic gravitational background from the Big Bang would lead to the decoherence  
of macroscopic quantities of matter even if other backgrounds were shielded. To maintain the 
quantum coherence of a system, it is necessary to shield the system, as much as possible, from all 
environmental backgrounds. If one considers the environmental effect coming from 
the cosmological photon, neutrino and gravity/graviton backgrounds associated with
the Big Bang, one should first consider the decohering effect of the strongest interacting background --
the photon Cosmic Microwave Background (CMB). However, electromagnetic radiation can be
easily shielded and this is generally the case when one is doing experiments
-- the apparatus of most experimental setups will tend to shield the material being tested
from the decohering effect of any background E\&M fields. The remaining backgrounds
from strongest to weakest are: (i) the neutrino background; (ii) the gravitational background.
It may seem strange to refer to the neutrino background as ``strongly interacting", 
but it is more strongly interacting than the gravitational background. This can be seen by comparing the
Fermi coupling $G_F \approx 1.17 \times 10^{-5} ~\rm{GeV}^{-2}$  
with Newton's constant $G_N \approx 6.71 \times 10^{-39} ~\rm{GeV}^{-2}$ 
both given in units of GeV$^{-2}$ (with $\hbar = c =1$) \cite{pdg}. 
In this paper we argue that the results of \cite{blencowe}, which point toward the
decoherence of macroscopic matter via gravitational backgrounds, can be repeated for a neutrino background.
Moreover the scale at which this neutrino decoherence occurs is at a much lower energy scale --
the weak scale rather than the Planck scale. This implies that it might be possible to observe this 
neutrino decoherence in individual nuclear systems. It is suggested that the effect of this
neutrino induced decoherence may be the cause for the observed variation of the decay rates of certain
nuclei \cite{fischbach}. In contrast the gravitational background only leads to observable decoherence effects for macroscopic 
aggregates of matter. 

\section{Decoherence via gravitational backgrounds}

First, we recap the general derivation for the cosmic gravitational background causing a rapid decoherence 
in macroscopic matter \cite{blencowe}. A more thorough and detailed discussion of decoherence can be found
in the review article \cite{zurek}. The Hamiltonian considered is that of a {\it system} coupled to a {\it bath}
of the form
\begin{equation}
\label{hamiltonian}
{\cal H} = \hbar \omega_0 a^{\dagger} a + 
\sum _i \left( \frac{p_i ^2}{2 m_i} + \frac{1}{2} m_i \omega _i ^2 q_i ^2 \right)
+ \hbar \omega_0 a^{\dagger} a \sum _i \lambda _i \frac{q_i}{\Delta _i}
\end{equation}
The first term, $\hbar \omega_0 a^{\dagger} a$, is the energy of the {\it system} with
natural frequency $\omega_0$. The second term, $\frac{p_i ^2}{2 m_i} + \frac{1}{2} m_i \omega _i ^2 q_i ^2$, 
is the energy (kinetic plus potential) of the {\it bath} where the index $i$ is the $i^{th}$ component
of the bath. The third term, $ \hbar \omega_0 a^{\dagger} a \sum _i \lambda _i \frac{q_i}{\Delta _i}$,
is the interaction between the system and the $i^{th}$ component of the bath with $\lambda _i$ being the
coupling strength and $\Delta _i =\sqrt{\hbar / (2 m_i \omega _i})$ is the zero point uncertainty of the bath's
$i^{th}$ component. Assuming an Ohmic bath spectral density of the form
\begin{equation}
\label{j}
\frac{J(\omega)}{(\hbar \omega _0)^2} = \pi \sum _i \lambda _i ^2 \delta (\omega - \omega_i )
\end{equation} 
one can show very generally \cite{blencowe} that, in the Born-Markov approximation, 
the decoherence rate of the system due to the interaction with the bath leads to 
a decoherence rate of 
\begin{equation}
\label{decohere}
\Gamma_{\lambda} \simeq \frac{k_B T}{\hbar} \left( \frac{E -{\tilde E}}{E_{\lambda}} \right)^2 ~,
\end{equation}
where $E- {\tilde E}$ is the energy 
difference between the ground state ${\tilde E}$ and the first excited state $E$, which is
associated with the frequency $\omega_0$; $T$ is the temperature of the bath; $E_{\lambda}$ is the energy scale 
set by the interaction strength(s), $\lambda_i$, between the bath and the system.
The linear dependence of the decoherence rate on temperature, $\Gamma_{\lambda} \propto T$, is worth 
remarking on since {\it a prior} one might have expected a different dependence on the temperature. 
In the review article on decoherence \cite{zurek} it is shown that, in general,
for systems like \eqref{hamiltonian} in the {\it high temperature limit} that the decoherence rate is linear
with respect to temperature {\it i.e.} $\propto T$. 

The decoherence rate in \eqref{decohere} depends on the system-bath interaction term in \eqref{hamiltonian} 
through the energy scale $E_\lambda$. The bath that gives the decoherence rate in \eqref{decohere} 
is a non-relativistic oscillator with masses $m_i$ and 
frequencies $\omega_i$. In \cite{blencowe} this non-relativistic oscillator bath was replaced by a relativistic, massless 
graviton bath with an action given by
\begin{equation}
\label{gravity-term}
S_{gravity} = \int d^4 x \left( -\frac{1}{2} \partial ^\rho h^{\mu \nu} \partial _\rho h_{\mu \nu}
+ \partial _\nu h^{\mu \nu} \partial ^\rho h_{\mu \rho} - 
\partial_\mu h \partial _\nu h^{\mu \nu} + \frac{1}{2} \partial ^\mu h \partial _\mu h \right) ~,
\end{equation}
where $h^{\mu \nu}$ is the deviation of the total metric from Minkowski, $\eta ^{\mu \nu}$ and $h = h_\mu ^\mu$. 
In \cite{blencowe} it was found that this relativistic, massless graviton bath led to 
the same decoherence rate, \eqref{decohere}, as the non-relativistic oscillator bath from \eqref{hamiltonian}.
Thus whether the bath is non-relativistic or relativistic does not appear to play a major role
in the general form of the decoherence rate \eqref{decohere}.  

Since the high temperature limit was used in arriving at the general decoherence rate in \eqref{decohere} so 
we make some comments on this assumption. From the review article \cite{zurek} the high temperature limit is 
defined as the condition that ``$k_B T$ is higher than all other relevant energy scales in the system". For cosmological neutrinos
with $T _\nu \sim 2$ K, which corresponds to an energy scale of $k_B T _\nu \sim 2 \times 10 ^{-4}$ eV, this temperature
is definitely large compared to systems which have milli-Kelvin temperatures or scalar fields at $T=0$ -- examples
which were considered in \cite{blencowe}. However, more relevant is how this temperature 
compares with the energy scale of the bath. For massless gravitons it was shown in \cite{blencowe} that the high 
temperature limit was justified. The case of the neutrino bath is complicated by the fact that neutrinos have
unknown, small masses. From direct measurements \cite{pdg} the upper limit on the neutrino mass is $m_\nu < 2$ eV.
If some flavor of neutrino had a mass just below this limit ({\it e.g.} $m_\nu =1$ eV) then it is clear that this 
mass-energy scale of the neutrino bath would exceed the bath temperature and the high temperature limit
would not be justified. From neutrino oscillation data \cite{pdg} the mass squared differences between 
neutrino generations 1 and 2, and between neutrino generations 2 and 3 are respectively
$\Delta m_{21} ^2 \sim 8 \times 10^{-5}$ eV$^2$ and $\Delta m_{32} ^2 \sim 2 \times 10^{-3}$ eV$^2$. From this
data it is not clear if one or more of the neutrinos ({\it i.e.} electron neutrino, muon 
neutrino, tau neutrino) has a mass smaller than the cosmic neutrino bath energy of $k_B T _\nu \sim 2 \times 10 ^{-4}$ eV,
which is needed to validate the use of the high temperature limit. If instead of cosmological neutrinos we consider solar
neutrinos as the bath, then the high temperature limit apparently would apply. Solar neutrinos come
predominately from the initial $pp$ reaction in the proton-proton chain -- $p + p \rightarrow d + e^+ + \nu _e$.
This reaction yields a continuous (but non-Planckian) neutrino spectrum with energies
running up to $\sim 0.4$ MeV. This energy  is certainly higher than the scale set by the neutrino masses or any other energy
scale in the system. However, while solar neutrinos from the $pp$ reaction have a continuous spectrum it is 
not a Planckian spectrum, and thus it is not clear how one would rigorously define the temperature which goes
into the decoherence rate formula \eqref{decohere}. Despite the solar neutrino spectrum not being 
strictly Planckian some decoherence rate similar to \eqref{decohere} should still apply. 

In \cite{blencowe} the gravitational background bath was coupled to a non-self-interacting, massive 
scalar field. The scalar field was taken as a model for bulk matter. In this case the strength of the interaction 
between the gravitational bath and the scalar field/bulk matter was given by the parameter 
$\lambda _i \rightarrow \kappa = \sqrt{32 \pi G_N}$ with $G_N$ being Newton's constant, and the energy 
scale, $E_\lambda$, was the Planck scale $E_{Pl} = \sqrt{\hbar c^2 /G_N}$. For this gravitational background the 
decoherence rate \eqref{decohere} became
\begin{equation}
\label{decohere-grav}
\Gamma_{gravity} \simeq \frac{k_B T}{\hbar} \left( \frac{E -{\tilde E}}{E_{Pl}} \right)^2 ~.
\end{equation} 
The decoherence time associated with this rate was $t_{gravity} = 1/\Gamma_{gravity}$. To get an estimate
of $t_{gravity}$ it was assumed, following \cite{kolb}, that the temperature of the graviton background was
$T_g \simeq 1 ~\rm{K}$. The Planck energy scale is given by $E_{Pl} \simeq 10^{19} ~\rm{GeV} \simeq 10^{-8} ~\rm{kg}$.
Taking MeV as our energy unit one can write Boltzmann's constant and the renormalized Planck constant as  
$k_B =8.62 \times 10 ^{-11} ~\rm{\frac{MeV}{K}}$ and $\hbar = 6.58 \times 10^{-22} ~\rm{MeV} \cdot \rm{sec}$ respectively. 
With this the pre-factor in \eqref{decohere-grav} is $\frac{k_B T}{\hbar} \simeq 10^{11} ~\rm{Hz}$. 
For {\it individual} atomic or nuclear systems with energy differences
of the order $E -{\tilde E} \simeq 1$ eV and $E -{\tilde E} \simeq 1$ MeV, respectively, one gets 
a small $\Gamma_{gravity}$ and large $t_{gravity}$. Explicitly for these generic atomic and nuclear energy differences one 
gets $\Gamma_{gravity} (atom) \simeq 10^{-45} ~\rm{Hz} ~;~ (t_{gravity} (atom) \simeq 10^{45} ~ \rm{sec})$ and
$\Gamma_{gravity} (nucleus) \simeq 10^{-33} ~\rm{Hz} ~;~(t_{gravity} (nucleus) \simeq 10^{33} ~ \rm{sec})$. 
Thus the decoherence effect of this gravitational background on micro-systems is negligible. However for macro-systems
with an Avogadro's number of atoms (i.e. $6.02 \times 10^{23}$ atoms), each having an energy difference of 
$E -{\tilde E} \simeq 1$ eV, one finds $\Gamma_{gravity} \simeq 10^{2} ~\rm{Hz} ~;~ (t_{gravity} \simeq 10^{-2} ~\rm{sec})$.
Thus for larger amounts of matter and/or nuclear energy scales this gravitational decoherence rate is faster and 
the decoherence time shorter. Therefore, in these macro-systems, gravitational decoherence is important. 
This was the main point of \cite{blencowe} -- stochastic gravitational backgrounds
could lead to decoherence of macroscopic collections of matter and
thus provide a mechanism for the classical nature of these macroscopic objects, even if all other backgrounds 
could be shielded. But for microscopic collections of matter, such as
individual atomic/nuclear systems, this gravitational decoherence would occur at time scales much longer than 
laboratory time scales so that gravitational decoherence would not be important for these microscopic systems. 

\section{Decoherence via neutrino backgrounds}

In this section we argue that the gravitational background decoherence of \cite{blencowe} should
carry over to the case where one replaces the stochastic gravitational background by a stochastic neutrino 
background such as the neutrino version of the CMB. Actually in the case of neutrinos there are other, more or less 
random sources one should consider -- the Sun, geo-neutrinos, neutrinos from nuclear reactors --
but here we focus on the neutrino version of the CMB. All the arguments of the previous section 
can be repeated but now using the effective field theory of the weak interaction -- the Fermi 
four-fermion interaction -- instead of the effective field theory of gravity \cite{eft-gravity, eft-gravity-2} 
used by Blencowe. In the generic system-bath Hamiltonian of \eqref{hamiltonian} the first term is
again some atomic or nuclear system characterized by some frequency $\omega _0$ which is 
associated with a system energy difference $E -{\tilde E}$. The next two terms in \eqref{hamiltonian}
are now the neutrino background/bath. The last term in \eqref{hamiltonian} is the interaction term between 
the system and the bath via the effective, four-fermion weak interaction. The coupling $\lambda _i$ now comes 
from the low energy effective field theory of the weak interaction. At low energies the weak interaction is characterized 
by the Fermi coupling $G_F \approx 1.17 \times 10^{-5} ~\rm{GeV}^{-2}$ \cite{pdg} (taking $\hbar = c = 1$). Thus 
for a bath interacting via the weak interaction one replaces Newton's constant $G_N \approx 6.71 \times 10^{-39} ~\rm{GeV}^{-2}$  
by the Fermi coupling $G_F$. In turn this means one replaces the
Planck energy scale in \eqref{decohere-grav} ({\it i.e.} $E_{Pl} = \sqrt{\frac{1}{G_N}} \approx 1.22 \times 10^{19}$ GeV)
by the weak energy scale $E_{weak} = \sqrt{\frac{1}{G_F}} \approx 300$ GeV. Thus the
neutrino induced decoherence rate of \eqref{decohere} becomes
\begin{equation}
\label{decohere-weak}
\Gamma_{weak} \simeq \frac{k_B T}{\hbar} \left( \frac{E -{\tilde E}}{E_{weak}} \right)^2 ~.
\end{equation}
Replacing $E_{weak}$ by $E_{gravity}$ means that the decoherence rate due to the
neutrino background will be $(E_{Pl} / E_{weak})^2 \simeq 10^{33}$ times larger 
and the decoherence time associated with the neutrino background will be $(E_{weak} / E_{Pl})^2 \simeq 10^{-33}$ 
times smaller, than the decoherence rate and time due to gravitational background. 

There are three elements that were important to decoherence via a stochastic  
gravitational background ground: (i) there is a gravitational equivalent to the CMB left over from the 
Big Bang with an estimated temperature of about $1 K$; (ii) all bulk matter interacts with gravity;
(iii) it is hard/impossible to shield matter from interacting with the gravitational background. We now look at
each of these elements in terms of a neutrino background.

In regard to the first point there is a well established theory of a cosmological neutrino background with an
estimated temperature of $T_{\nu} \simeq 2 ~\rm{K}$, which is comparable to the measured photon CMB 
temperature of $T_\gamma \simeq 2.7 ~\rm{K}$, and the estimated graviton CMB temperature of $T_g \simeq 1 ~\rm{K}$. 
Although neither the neutrino nor graviton CMB have been directly observed, due to the weakness of
both interactions, the case for the neutrino background is on firmer ground
than for the gravitational background. There is some indirect observational evidence 
\cite{neutrino} for the neutrino version of the CMB. Further there are other sources of neutrino backgrounds
(solar neutrinos, geo-neutrinos, reactor neutrinos) which may also give rise to neutrino decohering background.

In regard to the second point -- the universality of the
gravitational interaction with matter and especially bulk matter -- the neutrino background also interacts 
with all {\it matter} particle/fields {\it i.e.} quarks, electrons, muons, taus. Of all the
Standard Model particles it is only the photon and the gluon that the neutrino does not interact with directly. 
But the fact that the neutrino does not interact with photons and gluons is not important when discussing the 
decoherence of matter composed of quarks and leptons. What is important in terms of decoherence is the coupling
of neutrinos to the matter fields of the quarks and leptons -- or since quarks are permanently confined into protons and neutrons 
one wants the effective coupling of neutrinos to protons and neutrons. To investigate in more detail 
the coupling of neutrinos to matter fields we write down the weak charge $Q_{weak}$ \cite{pdg} for the 
up quark, down quarks and electron
\begin{equation}
\label{qweak}
Q^u _{weak} = 1 - \frac{8}{3} \sin ^2 \theta_W ~~~;~~~ Q^d _{weak} = -1 + \frac{4}{3} \sin ^2 \theta_W 
~~~;~~~ Q^e _{weak} = -1 + 4 \sin ^2 \theta_W  ~,
\end{equation}    
where $\theta _W$ is the Weinberg angle. In turn one can use the weak charge of the quarks to arrive at 
the weak charges for the proton and neutron 
\begin{equation}
\label{qweak-1}
Q^p _{weak} = 2 Q^u _{weak} +  Q^d _{weak} = 1 - 4 \sin ^2 \theta_W ~~~;~~~ 
Q^n _{weak} = Q^u _{weak} +  2 Q^d _{weak} = -1 ~.
\end{equation} 
Using the numerical value $sin ^2 \theta_W \approx 0.23$ \cite{pdg} the weak charge of the proton, neutron and electron is
\begin{equation}
\label{qweak-2}
Q^p _{weak} \approx +0.08~~~;~~~ Q^n _{weak} = -1  ~~~;~~~ Q^e _{weak}  \approx  -0.08~.
\end{equation}
This situation is almost the reverse of that for electric charge -- the electron and proton are
almost weak charge neutral, and have opposite weak charges, while the neutron has a substantial weak charge. 
Using the above notation the effective interaction Lagrangian between the neutrino and the electron, proton
and neutron respectively is \cite{quigg}
\begin{eqnarray}
\label{n-epn}
{\cal L}_{\nu - e} &=& \frac{G_F}{2 \sqrt{2}} 
[{\bar \nu} \gamma_\mu (1 - \gamma _5) \nu] \times [{\bar e} \gamma ^\mu (Q^e _{weak} - \gamma _5) e] ~, \nonumber \\
{\cal L}_{\nu - p} &=& \frac{G_F}{2 \sqrt{2}} 
[{\bar \nu} \gamma_\mu (1 - \gamma _5) \nu] \times [{\bar p} \gamma ^\mu (Q^p _{weak} - g_A \gamma _5) p] ~,\\
{\cal L}_{\nu - n} &=& \frac{G_F}{2 \sqrt{2}} 
[{\bar \nu} \gamma_\mu (1 - \gamma _5) \nu] \times [{\bar n} \gamma ^\mu (Q^n _{weak} - g_A \gamma _5) n] ~,\nonumber
\end{eqnarray} 
where $g_A = 1.254$ is the renormalized axial charge of the nucleon \cite{quigg} and $e, p, n$ are the electron, proton
and neutron spinors. In the non-relativistic limit of these matter fields the electron and nucleon amplitudes resulting 
from \eqref{n-epn} are \cite{quigg}
\begin{eqnarray}
\label{amp-epn}
{\cal M}_{\nu - e} &=& -i \frac{G_F Q^e _{weak}}{2 \sqrt{2}}  [{\bar \nu} \gamma_\mu (1 - \gamma _5) \nu] \rho _e (\bf r ) ~, \\
{\cal M}_{\nu - pn} &=& -i \frac{G_F Q^{pn} _{weak}}{2 \sqrt{2}} [{\bar \nu} \gamma_\mu (1 - \gamma _5) \nu] \rho _{pn} (\bf r) ~,
\end{eqnarray} 
where here $\rho _e (\bf r )$ is the electron density as a function of the neutrino coordinate ${\bf r}$ and
$Q^e _{weak} = - Z_e (1 - 4 \sin ^2 \theta_W)$ is the total weak charge of the atom with $Z_e$ being the number of
electrons in the atom. In the same way $\rho _{pn} (\bf r )$ is the nucleon density as a function of the neutrino 
coordinate ${\bf r}$ and $Q^{pn} _{weak} =  Z (1 - 4 \sin ^2 \theta_W) - N$ is the total weak charge of the nucleus
with $Z$ being the number of protons in the nucleus and $N$ being the numbers of neutrons in the nucleus. Making the
approximation that $1 - 4 \sin ^2 \theta_W \approx 0$ we find that the effective weak charge of some atom is approximately 
\begin{equation}
\label{eff-weak}
Q^{total} _{weak} \approx G_F N ~.
\end{equation}
The total weak charge of some atom is determined by the number of neutrons in the nucleus while the weak charge of the electron 
and proton drop out in this limit. Taking into account \eqref{eff-weak} the energy scale for the
decoherence rate becomes $\sqrt{\frac{1}{N G_F}}$ which in turn modifies the weak scale decoherence rate of 
\eqref{decohere-weak} as
\begin{equation}
\label{decohere-weak-2}
\Gamma_{weak} \simeq \frac{N k_B T}{\hbar} \left( \frac{E -{\tilde E}}{E_{weak}} \right)^2 ~.
\end{equation}
This indicates that the neutrino decoherence effect proposed here should be largest in atoms with neutron heavy nuclei. 
In the concluding section we will offer a potential experimental test for neutrino decoherence based on a variation 
of decay rates in radioactive nuclei which are neutron rich. 

There are many similarities between the gravitational decoherence of \cite{blencowe} and the neutrino decoherence
proposed here, but there is a difference in the topology of coupling. From \eqref{n-epn} one sees that
the effective weak coupling is a quartic coupling having four fields coming together at a space-time point -- two 
neutrino fields and two electron, proton or neutron fields. The gravitational coupling used in \cite{blencowe} was tri-linear 
in the the scalar and gravitational fields having the form
\begin{equation}
\label{g-phi}
{\cal L}_{g - \phi} = \frac{\kappa}{2}  T^{\mu \nu} (\phi ) h_{\mu \nu} ~, 
\end{equation} 
where $T^{\mu \nu} (\phi )$ is the energy-momentum tensor of the scalar field, $\phi$, which is quadratic
in the scalar field, and $h_{\mu \nu}$ is the linear gravitational field term. From a Feynman diagram point of
view \eqref{g-phi} represents two scalar fields connected with one graviton. One can ask if this tri-linear 
versus quartic coupling will invalidate using the decoherence rate from \eqref{decohere} for 
a neutrino background. The answer to this is ``no" since the effective weak couplings in \eqref{n-epn} 
are the low energy limits of the fundamental tri-linear couplings of the electroweak $Z^0$ boson to the 
electron, proton or neutron     
\begin{eqnarray}
\label{n-epn-2}
{\cal L}_{Z - e} &=& \frac{g}{4 \cos \theta _W } [{\bar e} \gamma ^\mu (Q^e _{weak} - \gamma _5) e] Z_\mu~, \nonumber \\
{\cal L}_{Z - p} &=& \frac{g}{4 \cos \theta _W } [{\bar p} \gamma ^\mu (Q^p _{weak} - g_A \gamma _5) p] Z_\mu~,\\
{\cal L}_{Z - n} &=& \frac{g}{4 \cos \theta _W } [{\bar n} \gamma ^\mu (Q^n _{weak} - g_A \gamma _5) n] Z_\mu~,\nonumber
\end{eqnarray}  
where $g$ is the fundamental $SU(2)$ coupling constant of the Standard Model and $Z_\mu$ is the field of the 
$Z^0$ boson. In comparing the fundamental electroweak coupling \eqref{n-epn-2} with gravitational coupling
\eqref{g-phi} we see that both are tri-linear -- for the gravitational case the scalar matter field is quadratic
via $T^{\mu \nu} (\phi )$ and linear via the ``gauge" boson $h_{\mu \nu}$, while for the electroweak case the 
spinor matter fields are quadratic through the square bracketed last terms in \eqref{n-epn-2} and the gauge boson
$Z_\mu$ appears linearly. This linear coupling of the $Z^0$ gauge boson is then in accord with the
linear coupling between system and bath in the general Hamiltonian of \eqref{hamiltonian}. 
Thus the fundamental form of the electroweak interaction as given in \eqref{n-epn-2}
has the same topology as the gravitational interaction given in \eqref{g-phi} -- the matter fields come in quadratically
and the gauge boson fields enter linearly. It is only in the low energy limit that the weak interaction appears
as quartic interaction. 

The third important ingredient for the proposed decoherence mechanism 
of \cite{blencowe} was the fact that gravitons, since they interact very weakly, are hard to shield.
In contrast CMB photons can be shielded. Even though neutrinos are more strongly interacting than 
gravitons, as can be seen by comparing $G_N$ with $G_F$, they are still much more weakly interacting than 
photons at the relevant energy scales. Because of this effective, weak interaction strength, neutrinos are
not easily shielded by ordinary matter -- they will pass through the walls of a nuclear reactor and even through
the entire Earth with a very low probability of scattering.

Since the weak scale is seventeen orders of magnitude smaller than the Planck scale this leads to a much larger 
decoherence rate for the neutrino background -- see \eqref{decohere-weak} and \eqref{decohere-weak-2} --
and thus a much shorter decoherence time. Thus unlike the decoherence effect from a stochastic 
gravitational background, which requires a macroscopic aggregate of matter in order to have
reasonable/measurable decoherence time, the decoherence due to a stochastic neutrino background
might be observable in individual nuclear (but not atomic) systems. To see this we repeat some of 
the numerical estimates for decoherence rates and times from the previous section but using 
\eqref{decohere-weak-2}. First, for a single atomic system with a typical energy difference
of $E -{\tilde E} \simeq 1$ eV \eqref{decohere-weak}, and again taking the temperature 
of the bath to be $T \approx 1 - 2$ K so that $\frac{k_B T}{\hbar} \simeq 10^{11} ~\rm{Hz}$, we find
\begin{equation}
\label{de-atom}
\Gamma_{weak (atomic)}  \simeq 10^{-12} Z_e (1 - 4 \sin ^2 \theta_W) ~\rm{Hz} \;\;\;  ; \; \;\;
t_{weak (atomic)}  \simeq \frac{10^{12}}{Z_e (1 - 4 \sin ^2 \theta_W)} ~\rm{sec}  ~,
\end{equation}
where we have taken into account the discussion around \eqref{amp-epn} where the effective weak coupling 
is modified by the weak charge of the atomic electrons as $G_F \rightarrow G_F Q^e _{weak}$. Even aside
from additional factor of $Q^e _{weak}$ one can see from \eqref{de-atom} that for atoms the decoherence rate is 
low and the decoherence times are long. In the approximation $\sin ^2 \theta_W \approx 1/4$ the decoherence
effect on atomic electrons would vanish. Turning now to individual nuclei with a typical energy difference of 
$E -{\tilde E} \simeq 1$ MeV we find 
\begin{equation}
\label{de-nuclei}
\Gamma_{weak (nuclear)} \simeq 2 | Z (1 - 4 \sin ^2 \theta_W) - N |~\rm{Hz} \;\;\; ; \;\;\; 
t_{weak (nuclear)}  \simeq \frac{0.5}{|Z (1 - 4 \sin ^2 \theta_W) - N|} ~\rm{sec} ~. 
\end{equation}
Even in the approximation $\sin ^2 \theta_W \approx 1/4$, where the proton contribution drops out, one sees from
\eqref{de-nuclei} that for individual nuclei the decoherence rate is high and the decoherence time is short. This
is especially true for neutron rich nuclei ({\it i.e.} for large $N$). 
The shorter decoherence time in \eqref{de-nuclei} comes from the fact that the weak energy scale, $E_{weak}$, is not so much larger
than the typical nuclear energy scale. This estimated neutrino background decoherence time for nuclear systems 
(especially neutron rich nuclei) indicates that, unlike the graviton decoherence, which only operates for macroscopic collections of
matter, the neutrino decoherence effect might be observable in individual nuclear systems. 

As a final comment we note that there are many more potential sources of background neutrinos as compared to
background gravitons, thus making neutrino decoherence (if it occurs) a much more common phenomenon. For gravitons there
are very few ``local" sources {\it e.g.} two inspiraling black holes. The only significant source of background gravitons 
is the gravitational version of the CMB. In contrast there are a host of sources for background neutrinos --
solar neutrinos from the Sun and terrestrial neutrinos from the core of the Earth or from nuclear reactors. 
On Earth the largest source of neutrinos comes from the $pp$ reaction in the Sun ($p + p \rightarrow d + e^+ + \nu _e$).
The neutrinos from this reaction have a continuous, but non-Planckian, spectrum up to an energy of about 
$0.4$ MeV. Thus for the case of solar neutrinos coming from the $pp$ reaction the high temperature limit,
that was used in arriving at the decoherence rate in \eqref{decohere}, is certainly valid. The only drawback
in the case of solar neutrinos from the $pp$ reaction is that since the spectrum is not strictly Planckian
one cannot rigorously define a temperature. 

\section{Further Discussion and Conclusions}

In this paper we have suggested that the mechanism of decoherence due a gravitational background 
presented in \cite{blencowe} will work even more effectively for a neutrino background. 
By considering a neutrino background, such as the neutrino version of the CMB, 
the weak scale, $E_{weak}$, replaces the Planck scale, $E_{Pl}$. Because of the seventeen 
orders of magnitude difference between the weak scale and Planck scale the estimate for the
weak decoherence rate from \eqref{decohere-weak} is 33 orders of magnitude larger than the gravitational
decoherence rate from \eqref{decohere-grav}. This in turn implies that the weak decoherence time is $10^{33}$ 
times shorter than the gravitational decoherence time. For nuclear energy scales ({\it i.e.} $E - {\tilde E} \simeq 1$ MeV)
one obtains weak decoherence times which are laboratory testable times for {\it individual} nuclear systems.
In contrast the gravitational decoherence only gave laboratory testable times for macroscopic collections of matter.
Except for the difference in scale -- $E_{Pl}$ versus $E_{weak}$ -- the gravitational  decoherence and the proposed
neutrino decoherence are similar: (i) Both are practically impossible to shield (in contrast it is fairly
easy to shield experiments from the photon CMB or other photon backgrounds); (ii) both gravitons and neutrinos 
couple to bulk matter (gravity couples to anything which has mass-energy and the neutrino couples to all matter
particles -- quarks and leptons -- via the weak interaction); (iii) there is a neutrino as well a graviton version of 
the photon CMB. In addition there are many more local sources of neutrino
background (the Sun, the Earth's interior, nuclear reactors) which could lead to neutrino decoherence effects. 

In regard to the possibility of experimentally seeing the neutrino decoherence effect on individual nuclear 
system one might think to use nuclear magnetic resonance methods (NMR) such as the qubits of quantum computing. 
This does not work for the following reason: for a typical qubit one has $E -{\tilde E} \simeq 10^{-3}$ eV 
\cite{q-computer} which from \eqref{decohere-weak} gives $t_{weak} = \frac{1}{\Gamma_{weak}} \simeq 10^{15} ~\rm{sec}$ 
which is much longer than the decoherence time for the qubits from other sources. The decoherence times for the qubit 
systems of \cite{q-computer} -- given by $T_2$ listed in figure 2 of \cite{q-computer} -- are on the order of $\sim 2$ 
seconds and thus dominate over $t_{weak}$. 

One speculative idea of potentially testing the decoherence effect of a neutrino background on nuclear systems 
involves the modification of nuclear decay rates via the quantum Zeno effect \cite{fonda} \cite{misra} {\it or}
the anti-quantum Zeno effect \cite{kofman}. It is known that the decay of an unstable 
quantum particle/system can be altered by making measurements of the system or by having the system interact with 
some environment \cite{watchdog} which causes decoherence of the system. In the quantum Zeno effect
the decay rate is reduced and thus the lifetime extended \cite{misra} while for the anti-quantum Zeno effect 
the decay rate is enhanced and thus the lifetime shortened \cite{kofman}. As detailed in \cite{kofman} whether 
some system exhibits the quantum Zeno effect or anti-quantum Zeno effect depends on various details of the system
and the bath. 
 
This neutrino decoherence mechanism for the variation of nuclear decay rates may have some experimental support in current unexplained 
deviations in nuclear decay rates. In reference \cite{fischbach} it was shown that the decay rates of $^{32} Si$ and
$^{226} Ra$ nuclei exhibited a small, but perceptible seasonal variation with respect to the orbit of the Earth around the Sun. 
One of the explanations offered for this variation was {\it some} unknown influence of the neutrinos emitted by the Sun
on the nuclei in question. The neutrino flux seen at the Earth changes yearly due to the changing distance of the 
Earth from the Sun and this would {\it somehow} lead to the yearly variation in the nuclear decay rates. 
The difficulty for this neutrino based explanation is that
one decay is a $\beta$-decay ($^{32}Si \rightarrow ^{32}P + \beta ^-  +{\bar \nu}_e$ releasing $\approx 0.2$ 
MeV of energy) and the other is an $\alpha$-decay ($^{226}Ra \rightarrow \alpha + ^{222}Rn$ releasing 
$\approx 4.9$ MeV of energy). It is hard to come up with some common particle physics explanation where changing
the neutrino flux would alter the decay rates of {\it both} $\alpha$-decay and $\beta$-decay. However neutrino induced 
decoherence would act on both $^{32} Si$ and $^{226} Ra$ -- both nuclei have a substantial number of neutrons and
thus from \eqref{qweak-1} have a substantial weak charge which would give a coupling between these nuclei and
the neutrino flux/bath. Also if one takes the energy difference ({\it i.e.} $E -{\tilde E}$) from 
\eqref{decohere-weak-2} to be the energy released in the decay -- $E -{\tilde E} \approx 0.2$ MeV for $^{32} Si$ decay
and $E -{\tilde E} \approx 4.9$ MeV for $^{226}Ra$ decay -- one finds large decoherence rates and short decoherence
times. For $^{32} Si$, using $E -{\tilde E} \simeq 0.2$ MeV, taking the temperature 
of the bath to be $T \approx 1 - 2$ K so that $\frac{k_B T}{\hbar} \simeq 10^{11} ~\rm{Hz}$, and
noting that the number of neutrons for $^{32} Si$ is $N=18$ equation \eqref{decohere-weak-2} gives 
\begin{equation}
\label{de-si32}
\Gamma_{weak (^{32} Si)}  \simeq 0.72 ~\rm{Hz} \;\;\; \;\; , \; \;\;\;
t_{weak (^{32} Si)}  \simeq 1.39 ~\rm{sec} ~.
\end{equation}
Similarly for $^{226} Ra$, which has $N=138$ equation \eqref{decohere-weak-2} gives
\begin{equation}
\label{de-ra226}
\Gamma_{weak (^{226} Ra)}  \simeq 3700 ~\rm{Hz} \;\;\; \;\; , \; \;\;\;
t_{weak (^{226} Ra)}  \simeq 2.7 \times 10 ^{-4} ~\rm{sec} ~.
\end{equation}
These decoherence rates for $^{32} Si$ and $^{226} Ra$ are large enough (and the decoherence times
are short enough) that the explanation of the variation of the decays rates via the quantum Zeno or anti-quantum Zeno
effect is possible. The above are estimates since the neutrino spectrum from the Sun is not a Planckian spectrum  
as is the case for the CMB neutrinos. Thus in the cases of $^{32} Si$ and $^{226} Ra$ one should calculate the decoherence 
rate using the changing non-Planckian flux of neutrinos coming from the Sun. The strong point of such an
explanation is that it is universal in the sense that it should occur whether or not the decays are $\alpha$, $\beta$,
or $\gamma$ decays. One potential test for this idea would be to look at nuclei that have some decay channel
whose energy is much lower than the usual nuclear scale of 1 MeV, say 1 keV. In such a case the estimated decoherence rate
would be much smaller and the decoherence time much larger so that one would not expect any variation of the
nuclear decay rate. Another, more extravagant test, would be to place radioactive material like
$^{32} Si$ and $^{226} Ra$ into a spacecraft with a highly elliptical orbit which would take it close to the Sun
at perihelion and then much farther from the Sun at aphelion. If neutrino decoherence is the cause of the
variation in the nuclear decay rates mentioned in \cite{fischbach} then the variation in nuclear decay rates 
should be much larger for the radioactive materials placed on such a spacecraft. 
Looking at the admittedly noisy data for the variation of the nuclear decay rates from 
\cite{fischbach} shows that the decay rates increase slightly to a maximum in January and decrease to a minimum
in July. Since the Earth is in perihelion in January and thus the neutrino flux should be greater then as compared
to aphelion which occurs in July, the above result indicates that it is the anti-quantum Zeno effect ({\it i.e.} increase
of the nuclear decay rate) which is operative. In subsequent work we will perform a more detailed analysis of 
the possibility of using the quantum Zeno versus anti-quantum Zeno effect to explain this variation of the nuclear decay rates
through decoherence. 
 
As a final comment we discuss why the neutrino decoherence effect goes as first order in $G_F$
(see for example \eqref{decohere-weak}, noting that $G_F \simeq \frac{1}{E_{weak} ^2}$)
rather than going as second order $G_F ^2$, which is the order of the weak scattering cross sections at low energy.
The same is true for gravitational decoherence of \cite{blencowe}-- it is proportional to $G_N$ rather than 
$G_N ^2$, the order of the gravitational scattering cross section. Thus the 
neutrino (and gravity) decoherence effect goes as the matrix element rather than the matrix element squared. 
The explanation for this is that the proposed decoherence effect can be viewed as the
inverse or complement of the Mikheev-Smirnov-Wolfenstein (MSW) effect \cite{MSW, MSW2}. In the MSW effect the oscillations 
of the quantum phases of neutrinos are altered when the neutrinos travel through bulk
matter. The MSW effect depends on the forward scattering {\it matrix element} of the weak interaction between the 
neutrinos and the background matter ({\it i.e.} electrons, protons and neutrons)  through which the neutrinos are 
traveling. Even vacuum neutrino oscillations can be viewed as a decoherence effect 
\cite{ohlsson}. Here we are proposing that the quantum phase coherence of matter  
is altered by its interaction with the neutrino background in which the matter is embedded. 
Identifying the proposed neutrino decoherence effect as the complement or inverse of the
MSW effect shows why the neutrino decoherence is of order $G_F$ rather than order $G_F ^2$ -- 
the regular MSW effect is proportional to the forward scattering matrix element which goes as $G_F$. 
In contrast the scattering cross section is related to the matrix element squared and is thus of order $G_F ^2$.
To bring the discussion back full circle to gravitational decoherence we note that in the gravitational context there is
also a similar MSW-like effect as discussed by several authors \cite{ahluwalia1} \cite{ahluwalia2} \cite{konno}
\cite{wudka}

\end{document}